\documentclass[12pt]{revtex4-1}
\usepackage{amsmath}
\usepackage{amssymb}
\usepackage[left=1cm,right=1cm]{geometry}
\usepackage[colorlinks,linkcolor=blue,citecolor=red]{hyperref}
\usepackage{feynmp}
\usepackage{slashed}
\usepackage{color}
\usepackage{graphicx}
\usepackage{makecell}
\usepackage{multirow}
\usepackage{braket}

\begin{document}

\title{\LARGE Path Integral Method for Pricing Proportional Step Double-Barrier Option with Time Dependent Parameters}
\bigskip
\author{Qi Chen~$^{1}$}
\email{angel.chern0214@gmail.com}
\author{Chao Guo~$^{2}$}
\email{chaog@lfnu.edu.cn}
\affiliation{
$^{1}$~School of Economics and Management, Langfang Normal University, Langfang 065000, China
\\
$^2$ School of Science, Langfang Normal University, Langfang 065000, China
}
\date{\today}

\begin{abstract}
Path integral method in quantum mechanics provides a new thinking for barrier option pricing. For proportional double-barrier step (PDBS) options, the option price changing process is analogous to a particle moving in a finite symmetric square potential well. We have derived the pricing kernel of PDBS options with time dependent interest rate and volatility. Numerical results of option price as a function of underlying asset price are shown as well. Path integral method can be easily generalized to the pricing of PDBS options with curved boundaries.
\end{abstract}

\maketitle

\section{Introduction}

Black-Scholes (BS) model gives a mathematical method for European option pricing by solving stochastic partial differential equations (SPDEs)~\cite{Black}. Using a variable substitution, the BS equation could be rewritten into a Schr{\"o}dinger type equation~\cite{Baaquie:2004}. For European options with a fixed interest rate and volatility, the option price changing with time could be analogous to the evolution of a one-dimensional free particle in the space~\cite{Baaquie:2004}. For a double-barrier option, the option price changing with time could be analogous to a particle moving in a infinite square potential well~\cite{Baaquie:2004,Baaquie}. Recently, a series of barrier options have emerged in financial markets. Linetsky discussed a kind of option: when the underlying price touches and passes the barrier, the option contract is not invalid immediately, but the option knocks out (knocks in) gradually, which is called step option~\cite{Linetsky:1999,Linetsky:2001}. A proportional step call is defined by its payoff~\cite{Linetsky:1999}
\begin{equation}\label{step def}
     e^{-V_0\tau} {\rm max}(S_T-K,0)
 \end{equation}
 where $S_T$ is the underlying asset price on the expiration date, $K$ is the exercise price, $\tau$ is the time to maturity. Chen et al.~\cite{Chen} provides a path integral method to calculate the proportional step call and the proportional double-barrier step call price, where $V_0$ in (\ref{step def}) is viewed as the depth of the well of one-dimensional problem in quantum mechanics. The pricing methods of barrier options with time dependent parameters are considered in a series of papers. Barrier options with curved boundaries were first introduced in~\cite{Kunitomo:1992}. Using path integral method, Chen et al.~\cite{Chen} discussed the pricing of such options and derived the option price analytical expressions. Lo et al.~\cite{Lo:2003} discussed the pricing approach of barrier options with time dependent interest rate and volatility by solving the BS equation. Besides the time dependent interest rate and volatility, Roberts and Shortland~\cite{Roberts:1997} also considered options with time dependent barrier levels, and computed the barrier option prices. Alghalith~\cite{Alghalith:2020,Alghalith:2021} provided a closed form formula for the pricing of European options with stochastic volatility and interest rate, and this approach could be generalized to price barrier options. However, step options with time dependent parameters has not been discussed. In this paper, we focus on deriving the analytical expressions of the proportional double-barrier step (PDBS) option price by path integral method, and give the numerical results.  

 Our work is organized as follows. In Section 2, we will derive the analytical expressions for the PDBS option with time dependent interest rate and volatility by path integral method. In Section 3,  we show the numerical results for option price as a function of underlying asset price. We summarize our main results in Section 4. The pricing formulas for the standard proportional double-barrier step call option derived by path integral method is reviewed in Appendix A.
\section{Proportional Step Double-Barrier Option with Time Dependent Interest Rate}

The price changing of a proportional double-barrier step ({\rm PDBS}) option could be analogous to a particle moving in a symmetric square potential well with the potential
\begin{equation}\label{potential1}
V(x)=\left\{
\begin{aligned}
0 & , & a<x<b,\\
V_0 & , & x<a,\ x>b.
\end{aligned}
\right.
\end{equation}
for $x<a$ or $x>b$, the wave function decays with the increasing distances from the well, which is similar to an option touches a barrier and knocks out gradually. The Hamitonian for a double-barrier step option derived from Black-Scholes equation is~\cite{Baaquie}
\begin{equation}\label{nonhermitehami}
H_{\rm PDBS}=-\frac{\sigma^2(t)}{2}\frac{\partial^2}{\partial x^2}+\left(\frac{1}{2}\sigma^2(t)-r(t)\right)\frac{\partial}{\partial x}+r(t)+V(x)
\end{equation}
where the interest rate $r(t)$ and the volatility $\sigma(t)$ change with time, and (\ref{nonhermitehami}) is a non-Hermitian Hamitonian. Considering the following transformation
\begin{equation}\label{hermitehami}
H_{\rm PDBS}=e^{\alpha(t) x} H_{\rm {eff}}e^{-\alpha(t) x}=
e^{\alpha(t) x}\left(-\frac{\sigma^2(t)}{2}\frac{\partial^2}{\partial x^2}+\gamma(t)\right)e^{-\alpha(t) x}+V(x)
\end{equation} 
where
\begin{equation}
\alpha(t)=\frac{1}{\sigma^2(t)}\left(\frac{\sigma^2}{2}-r(t)\right),\\ 
\gamma(t)=\frac{1}{2\sigma^2(t)}\left(\frac{\sigma^2}{2}+r(t)\right)^2
\end{equation}
and $H_{\rm {eff}}$ is a Hermitian Hamitonian which is considered as the symmetric square potential well Hamitonian. The stationary state Schr{\"o}dinger equation for option price is
\begin{equation}
    \left\{
\begin{aligned}\label{schrodingereq}
&-\frac{\sigma^2}{2}\frac{{\rm d}^2 \phi}{{\rm d}x^2}+\gamma(t) \phi=E\phi  , & a<x<b,\\
&-\frac{\sigma^2}{2}\frac{{\rm d}^2 \phi}{{\rm d}x^2}+(\gamma(t)+V_0) \phi=E\phi  , & x<a,\ x>b.
\end{aligned}
\right.
\end{equation}
where $\phi$ is the option price wave function, $E$ is corresponding to bound state energy levels in the potential well. (\ref{schrodingereq}) could be simplified into
\begin{equation}
    \left\{
\begin{aligned}\label{simpleschrodinger1}
\frac{{\rm d}^2 \phi}{{\rm d}x^2}+k_1^2 \phi=0 & , & a<x<b,\\
\frac{{\rm d}^2 \phi}{{\rm d}x^2}-k_2^2 \phi=0 & , & x<a,\ x>b.
\end{aligned}
\right.
\end{equation}
with the definition 
\begin{equation}\label{k1k2}
    k_1^2=\frac{2(E-\gamma(t))}{\sigma^2},\ \ k_2^2=\frac{2(V_0+\gamma(t)-E)}{\sigma^2}
\end{equation}

The general solution for (\ref{simpleschrodinger1}) is
\begin{equation}\label{general solution}
\phi(x)=\left\{
\begin{aligned}
A_3\ e^{k_2(x-\frac{b+a}{2})} & , & x\leq a, \\
A_1 \sin(k_1x+\delta) & , & a< x \leq b, \\
A_2\ e^{-k_2(x-\frac{b+a}{2})} & , & x>b.
\end{aligned}
\right.
\end{equation}
considering the continuity for both wave function and its derivative at $x=a$ and $x=b$, we have 
\begin{equation}\label{delta value}
    \delta=\frac{\ell\pi}{2}-k_1\frac{b+a}{2},\ \ \ell=0, 1, 2,...
\end{equation}

According to different $\ell s$ in (\ref{delta value}), (\ref{general solution}) could be split into two parts

\begin{equation}\label{oddeq}
\phi_1(x)=\left\{
\begin{aligned}
&-A_{21}e^{k_2(x-\frac{b+a}{2})}  , & x\leq a, \\
&A_1 \sin k_1\left(x-\frac{b+a}{2}\right) , & a\leq x \leq b,\\ 
&A_{21} e^{-k_2(x-\frac{b+a}{2})}  , & x>b.
\end{aligned}
\right.\ \ \ \ for\ \ \ell=0, 2, 4,...
\end{equation}
and
\begin{equation}\label{eveneq}
\phi_2(x)=\left\{
\begin{aligned}
&A_{22} e^{k_2(x-\frac{b+a}{2})}  , & x\leq a, \\
&A_1 \cos k_1\left(x-\frac{b+a}{2}\right)  , & a\leq x \leq b, \\
&A_{22} e^{-k_2(x-\frac{b+a}{2})}  , & x>b.
\end{aligned}
\right.\ \ \ \ for\ \ \ell=1, 3, 5,...
\end{equation}	
where 
\begin{equation}\label{A1A2}
    A_1=\sqrt{\frac{2k_2}{k_2(b-a)+2}},\ \ A_{21}=A_1\sin \frac{k_1(b-a)}{2}e^{k_2\frac{b-a}{2}},\ \ A_{22}=A_1\cos \frac{k_1(b-a)}{2}e^{k_2\frac{b-a}{2}}
\end{equation}
here the normalization condition has been used. Considering boundary conditions for (\ref{oddeq}) and (\ref{eveneq}) at $x=b$ respectively, we have
\begin{equation}\label{condition1}
    \cot k_1\frac{b-a}{2}=-\frac{k_2}{k_1}
    \end{equation}
    \begin{equation}\label{condition2}
    \tan k_1\frac{b-a}{2}=\frac{k_2}{k_1}
\end{equation}
let 
\begin{equation}\label{beta}
     \beta=\sqrt{k_1^2+k_2^2}=\frac{\sqrt{2V_0}}{\sigma(t)}
\end{equation}
(\ref{condition1}) and (\ref{condition2}) could be combined into
\begin{equation}\label{conditionk1}
    k_{1n}\frac{b-a}{2}=\frac{n\pi}{2}-\arcsin{\frac{k_{1n}}{\beta}},\ \ n=1,2,3,...
\end{equation}
(\ref{conditionk1}) is the energy level equation for the option. In general, there is no accurate analytical solution for energy eigenvalues, for low energy levels ($k_{1n}\ll\beta$), $\arcsin{({k_{1n}}/\beta)}\approx {k_{1n}}/\beta$, the approximate expression for $k_{1n}$ from (\ref{conditionk1}) is
\begin{equation}
    k_{1n,{\rm low}}\approx \frac{\beta n\pi}{\beta(b-a)+2}
\end{equation}
which is dependent on $t$. The contribution to the option price from high energy levels $(k_{1n}\approx\beta)$ is negligible~\cite{Chen}, we only take low energy levels into account in the following discussions. 

Now we calculate the expressions of option pricing kernels, which could be written as
\begin{equation}\label{pricingkernelPDBS}
    P_{\rm PDBS}(x,x^\prime;\epsilon,\tau)=\braket{x|e^{-\tau_1H_1-\tau_2H_2}|x^\prime}
    =\int_{-\infty}^{+\infty}{\rm d}x^{\prime\prime}\braket{x|e^{-\tau_1H_1}|x^{\prime\prime}}\braket{x^{\prime\prime}|e^{-\tau_2H_2}|x^{\prime}}
\end{equation}
where $\tau_1$ indicates the occupation time between the lower barrier a and the upper barrier b, and $\tau_2$ is the occupation time below the lower barrier a and above the upper barrier b, and
\begin{equation}\begin{split}
H_1&=
e^{\alpha(t) x}\left(-\frac{\sigma^2(t)}{2}\frac{\partial^2}{\partial x^2}+\gamma(t)\right)e^{-\alpha(t) x}\\
H_2&=
e^{\alpha(t) x}\left(-\frac{\sigma^2(t)}{2}\frac{\partial^2}{\partial x^2}+\gamma(t)\right)e^{-\alpha(t) x}+V_0
\end{split}\end{equation}
since $\gamma(t)$ is dependent on $t$,  we discretize $\tau$ so that there are $N$ steps to maturity, with each time step is  
\begin{equation}\label{eq:tinytime}
    \epsilon=\tau/N
\end{equation}
and 
\begin{equation}
\tau_1=M\epsilon,\ \ \ \tau_2=\tau-\tau_1=(N-M)\epsilon
\end{equation}
the first propagator in (\ref{pricingkernelPDBS}) could be denoted as
\begin{equation}\begin{split}\label{propagator}
\braket{x|e^{-\tau_1H_1}|x^{\prime\prime}}&=\int_{-\infty}^{+\infty}{\rm d}x_1\int_{-\infty}^{+\infty}{\rm d}x_2...\int_{\infty}^{\infty}{\rm d}x_j...\int_{-\infty}^{+\infty}{\rm d}x_{M-1}\times\\
&\braket{x|e^{-\epsilon H_1}|x_1}\braket{x_1|e^{-\epsilon H_1}|x_2}...\braket{x_{j-1}|e^{-\epsilon H_1}|x_j}...\braket{x_{M-1}|e^{-\epsilon H_1}|x^{\prime\prime}}
\end{split}\end{equation}
$jth$ matrix element in (\ref{propagator}) is
\begin{equation}\begin{split}
\braket{x_{j-1}|e^{-\epsilon H_1}|x_j}&=\sum_n\braket{x_{j-1}|e^{-\epsilon H_1}|n}\braket{n|x_j}\\
&=e^{\alpha_{j-1}x_{j-1}-\alpha_j x_j}e^{-\epsilon\gamma_{j-1}}\sum_n e^{-\frac{1}{2}\epsilon\sigma_{j-1}^2k_{1n,j-1}^2}\phi_n(x_{j-1})\phi_n({x_j})
\end{split}\end{equation}
where $\phi_n(x)$ is given by (\ref{oddeq}) and (\ref{eveneq}).
Similarly
\begin{equation}\begin{split}
    \braket{x_{j-1}|e^{-2\epsilon H_{\rm DFB}}|x_{j+1}}&=\int_{-\infty}^{+\infty}{\rm d}x_j \braket{x_{j-1}|e^{-\epsilon H_{\rm DFB}}|x_j}\braket{x_j|e^{-\epsilon H_{\rm DFB}}|x_{j+1}}\\
    &=\int_{-\infty}^{+\infty}{\rm d}x_j e^{\alpha_{j-1}x_{j-1}-\alpha_{j+1}x_{j+1}}e^{-\epsilon(\gamma_{j-1}+\gamma_j)}\sum_{n}\sum_{n^\prime}e^{-\frac{1}{2}\epsilon(\sigma_{j-1}^2k_{1n,j-1}^2+\sigma_j^2k_{1n^\prime,j}^2)}\times\\
    &\phi_n(x_{j-1})\phi_n(x_{j})\phi_{n^\prime}(x_j)\phi_{n^\prime}(x_{j+1})\\
    &=e^{\alpha_{j-1}x_{j-1}-\alpha_{j+1}x_{j+1}}e^{-\epsilon(\gamma_{j-1}+\gamma_j)}\sum_n e^{-\frac{1}{2}\epsilon(\sigma_{j-1}^2k_{1n,j-1}^2+\sigma_j^2k_{1n^\prime,j}^2)}\phi_n(x_{j-1})\phi_n(x_{j+1})
\end{split}\end{equation}
where the orthonormalization condition
\begin{equation}
    \int_{-\infty}^{+\infty}{\rm d}x_j\phi_n(x_j)\phi_{n^\prime}(x_j)=\delta_{nn^\prime}=\left\{
\begin{aligned}
0 & , & n \neq n^\prime,\\
1 & , & n=n^\prime .
\end{aligned}
\right.
\end{equation}
has been used. After some similar calculation, the propagator (\ref{propagator}) is
\begin{equation}\label{firstpropagator}
\braket{x|e^{-\tau_1H_1}|x^{\prime\prime}}=e^{\alpha_0 x-\alpha_M x^{\prime\prime}}e^{-\lim\limits_{\epsilon\to 0}\epsilon\sum\limits_{j=0}\limits^{M-1}\gamma_j}\sum_n e^{-\frac{1}{2}\lim\limits_{\epsilon\to 0}\epsilon\sum\limits_{j=0}\limits^{M-1}\sigma_j^2 k_{1n,j}^2}\phi_n(x)\phi_n(x^{\prime\prime})
\end{equation}
the second propagator in (\ref{pricingkernelPDBS}) could be calculated as
\begin{equation}\label{secondpropagator}
\braket{x^{\prime\prime}|e^{-\tau_2H_2}|x^\prime}=e^{\alpha_M x^{\prime\prime}-\alpha_N x^{\prime}}e^{-\lim\limits_{\epsilon\to 0}\epsilon\sum\limits_{j=M}\limits^{N-1}\gamma_j}\sum_{n^\prime} e^{-\frac{1}{2}\lim\limits_{\epsilon\to 0}\epsilon\sum\limits_{j=M}\limits^{N-1}\sigma_j^2 k_{{2n^\prime},j}^2}\phi_{n^\prime}(x^{\prime\prime})\phi_{n^\prime}(x^\prime)
\end{equation}
and the pricing kernel (\ref{pricingkernelPDBS}) is
\begin{equation}\label{kerneleq}
P_{\rm PDBS}(x,x^\prime;\epsilon,\tau)=e^{\alpha_0 x-\alpha_N x^{\prime\prime}}e^{-\lim\limits_{\epsilon\to 0}\epsilon\sum\limits_{j=0}\limits^{N-1}\gamma_j}\sum_{n} e^{-\frac{1}{2}\lim\limits_{\epsilon\to 0}\epsilon\sum\limits_{j=0}\limits^{N-1}\sigma_j^2 k_{{1n},j}^2}\phi_{n}(x)\phi_{n}(x^\prime)
\end{equation}
the option price could be denoted as
\begin{equation}
C_{\rm PDBS}(x,x^\prime;\tau)=\int_{{\rm ln}K}^{+\infty}P_{\rm PDBS}(x,x^\prime;\epsilon,\tau)(e^{x^\prime}-K)
\end{equation}
where $K$ is the exercise price. We will use pricing kernel (\ref{kerneleq}) to discuss the following three different cases:
\begin{itemize}
    \item $r(t)=0.05+0.01t$, \ \ $\sigma=0.3$
\end{itemize}
the series in (\ref{kerneleq}) could be denoted as
\begin{equation}\begin{split}
\lim_{\epsilon\to 0}\epsilon\sum_{j=0}^{N-1}\gamma_j&=\lim_{\epsilon\to 0}\epsilon\sum_{j=0}^{N-1}\frac{1}{2\sigma^2}\bigg(\frac{\sigma^2}{2}+0.05+0.01j\epsilon\bigg)^2\\
&=\frac{1}{0.18}\lim_{\epsilon\to 0 }\bigg(0.095^2\epsilon N+0.019\epsilon^2\sum_{j=0}^{N-1}j+0.0001\epsilon^3\sum_{j=0}^{N-1}j^2\bigg)\\
&=\frac{1}{0.18}\bigg(0.095^2\tau+0.019\times\frac{\tau^2}{2}+0.0001\times\frac{\tau^3}{3}\bigg)
\end{split}\end{equation}
\begin{equation}
\lim_{\epsilon\to 0}\epsilon\sum_{j=0}^{N-1}\sigma_j^2 k_{1n,j}^2=\sigma^2 k_{1n}^2\epsilon N=\sigma^2\tau\frac{\beta^2 n^2\pi^2}{[\beta(b-a)+2]^2}=\frac{0.18n^2\pi^2 V_0\tau}{[\sqrt{2V_0}(b-a)+0.6]^2}
\end{equation}
\begin{itemize}
    \item $r(t)=0.04+0.01e^{-t}$, \ \ $\sigma=0.3$
\end{itemize}
\begin{equation}\begin{split}
\lim_{\epsilon\to 0}\epsilon\sum_{j=0}^{N-1}\gamma_j&=\lim_{\epsilon\to 0}\epsilon\sum_{j=0}^{N-1}\frac{1}{2\sigma^2}\bigg(\frac{\sigma^2}{2}+0.04+0.01e^{-j\epsilon}\bigg)^2\\
&=\frac{1}{0.18}\lim_{\epsilon\to 0}\bigg(0.0085^2\epsilon N+0.0017\epsilon\sum_{j=0}^{N-1}e^{-j\epsilon}+0.0001\epsilon\sum_{j=0}^{N-1}e^{-2j\epsilon}\bigg)\\
&=\frac{1}{0.18}\bigg[0.0085^2\tau+0.0017\lim_{\epsilon\to 0}\frac{\epsilon(1-e^{-\tau})}{1-e^{-\epsilon}}+0.0001\lim_{\epsilon\to 0}\frac{\epsilon(1-e^{-2\tau})}{1-e^{-2\epsilon}}\bigg]\\
&=\frac{1}{0.18}\bigg[0.085^2+0.0017(1-e^{-\tau})+0.00005(1-e^{-2\tau})\bigg]
\end{split}\end{equation}
\begin{equation}
\lim_{\epsilon\to 0}\epsilon\sum_{j=0}^{N-1}\sigma_j^2 k_{1n,j}^2=\frac{0.18n^2\pi^2 V_0\tau}{[\sqrt{2V_0}(b-a)+0.6]^2}
\end{equation}
\begin{itemize}
    \item $r(t)=0.05$, \ \ $\sigma=0.3+0.1t$
\end{itemize}
\begin{equation}\begin{split}
\lim_{\epsilon\to 0}\epsilon\sum_{j=0}^{N-1}\gamma_j&=\lim_{\epsilon\to 0}\sum_{j=0}^{N-1}\frac{\epsilon}{2(0.3+0.1j\epsilon)^2}\bigg[\frac{(0.3+0.1j\epsilon)^2}{2}-0.05\bigg]^2\\
&=\lim_{\epsilon\to 0}\bigg[\frac{1}{8}\bigg(0.09\epsilon N+0.03\epsilon^2\sum_{j=0}^{N-1}j+0.0025\epsilon^3\sum_{j=0}^{N-1}j^2\bigg)+0.025\epsilon N+\frac{1}{2\epsilon}\sum_{j=0}^{N-1}\frac{1}{(j+\frac{6}{\epsilon})^2}\bigg]\\
&=\frac{1}{8}\bigg(0.09\tau+0.03\times\frac{\tau^2}{2}+0.0025\times\frac{\tau^3}{3}\bigg)+0.0025\tau+\lim_{\epsilon\to 0}\frac{1}{2\epsilon}\bigg[\psi^\prime\bigg(\frac{6}{\epsilon}\bigg)-\psi^\prime\bigg(\frac{6}{\epsilon}+N\bigg)\bigg]
\end{split}\end{equation}
\begin{equation}\begin{split}
\epsilon\sum_{j=0}^{N-1}\sigma_j^2 k_{1n,j}^2&=\epsilon\sum_{j=0}^{N-1}(0.3+0.05j\epsilon)^2\frac{\beta^2n^2\pi^2}{[\beta(b-a)+2]^2}\\
&=\frac{1}{2}n^2\pi^2V_0\epsilon N-\frac{n^2\pi^2(b-a)V_0\sqrt{2V_0}}{0.1}\sum_{j=0}^{N-1}\frac{1}{j+\frac{1}{\epsilon}\big(6+\frac{\sqrt{2V_0}(b-a)}{0.1}\big)}\\
&+\frac{n^2\pi^2(b-a)^2V_0^2}{0.01\epsilon}\sum_{j=0}^{N-1}\frac{1}{\big[j+\frac{1}{\epsilon}\big(6+\frac{\sqrt{2V_0}(b-a)}{0.1}\big)\big]^2}\\
&=\frac{1}{2}n^2\pi^2V_0\tau-\frac{n^2\pi^2(b-a)V_0\sqrt{2V_0}}{0.1}\times\\
&\bigg[\psi\bigg(\frac{1}{\epsilon}\bigg(6+\frac{\sqrt{2V_0}(b-a)}{0.1}\bigg)\bigg)-\psi\bigg(\frac{1}{\epsilon}\bigg(6+\frac{\sqrt{2V_0}(b-a)}{0.1}\bigg)+N\bigg)\bigg]+\\
&\frac{n^2\pi^2(b-a)^2V_0^2}{0.01\epsilon}\bigg[\psi^{\prime}\bigg(\frac{1}{\epsilon}\bigg(6+\frac{\sqrt{2V_0}(b-a)}{0.1}\bigg)\bigg)-\psi^{\prime}\bigg(\frac{1}{\epsilon}\bigg(6+\frac{\sqrt{2V_0}(b-a)}{0.1}\bigg)+N\bigg)\bigg]
\end{split}\end{equation}
where 
\begin{equation}
    \psi(z)=\frac{\Gamma^\prime(z)}{\Gamma(z)}
\end{equation}
is the $\psi$ function, and
\begin{equation}
    \Gamma(z)=\int_0^\infty e^{-t}t^{z-1}{\rm d}t,\ \ {\rm Re}z>0
\end{equation}
is the $\Gamma$ function.
\section{Numerical Results}
\begin{figure}[!htbp]
\begin{center}
\includegraphics[width=0.49\linewidth]{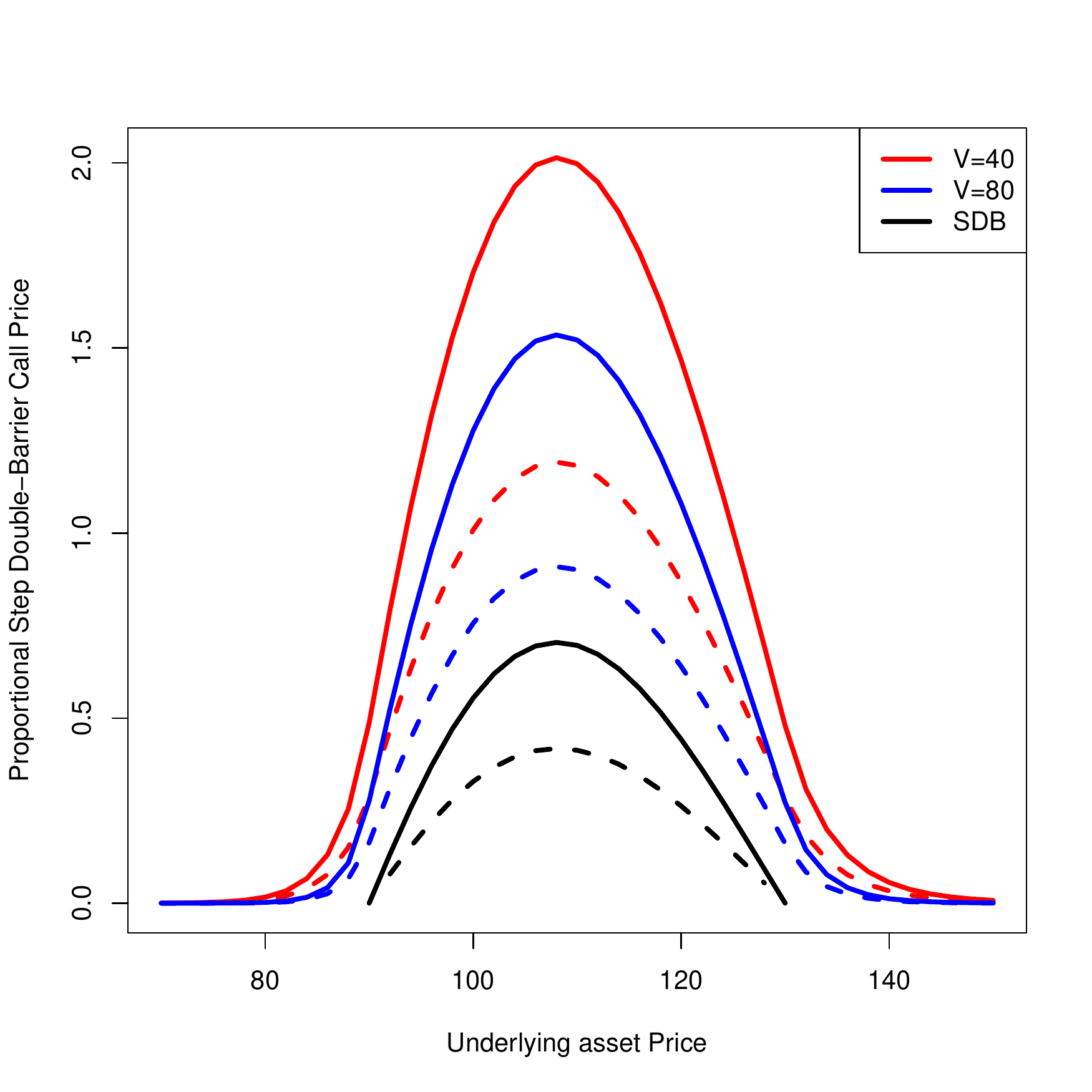}
\end{center}
\caption{Proportional double-barrier step call price as a function of  the underlying asset price with time-dependent interest rate $r(t)=0.05+0.01t$ for different  potentials. The dashed lines indicate the corresponding fixed interest rate $r=0.05$ cases for comparison. Parameters: $a={\rm ln}100=4.5$, $b={\rm ln}130=4.867$, $K=100$, $\sigma=0.3$, $\tau=1$.}
\label{fig: Pricert}
\end{figure}

In Fig.~\ref{fig: Pricert},  we show the proportional double-barrier step call price  as a function of underlying price, where the time-dependent interest rate is $r(t)=0.05+0.01t$. The dashed lines are corresponding to fixed interest rate $r=0.05$ cases for comparison. It is shown that the option prices decrease with the increasing of potential $V_0$, and the option prices with an interest rate increasing over time is higher than the one under fixed interest rate. In the limit $V_0\to\infty$, the option payoff tends to be the payoff of a standard double-barrier (SDB) option.

\begin{figure}[!htbp]
\begin{center}
\includegraphics[width=0.49\linewidth]{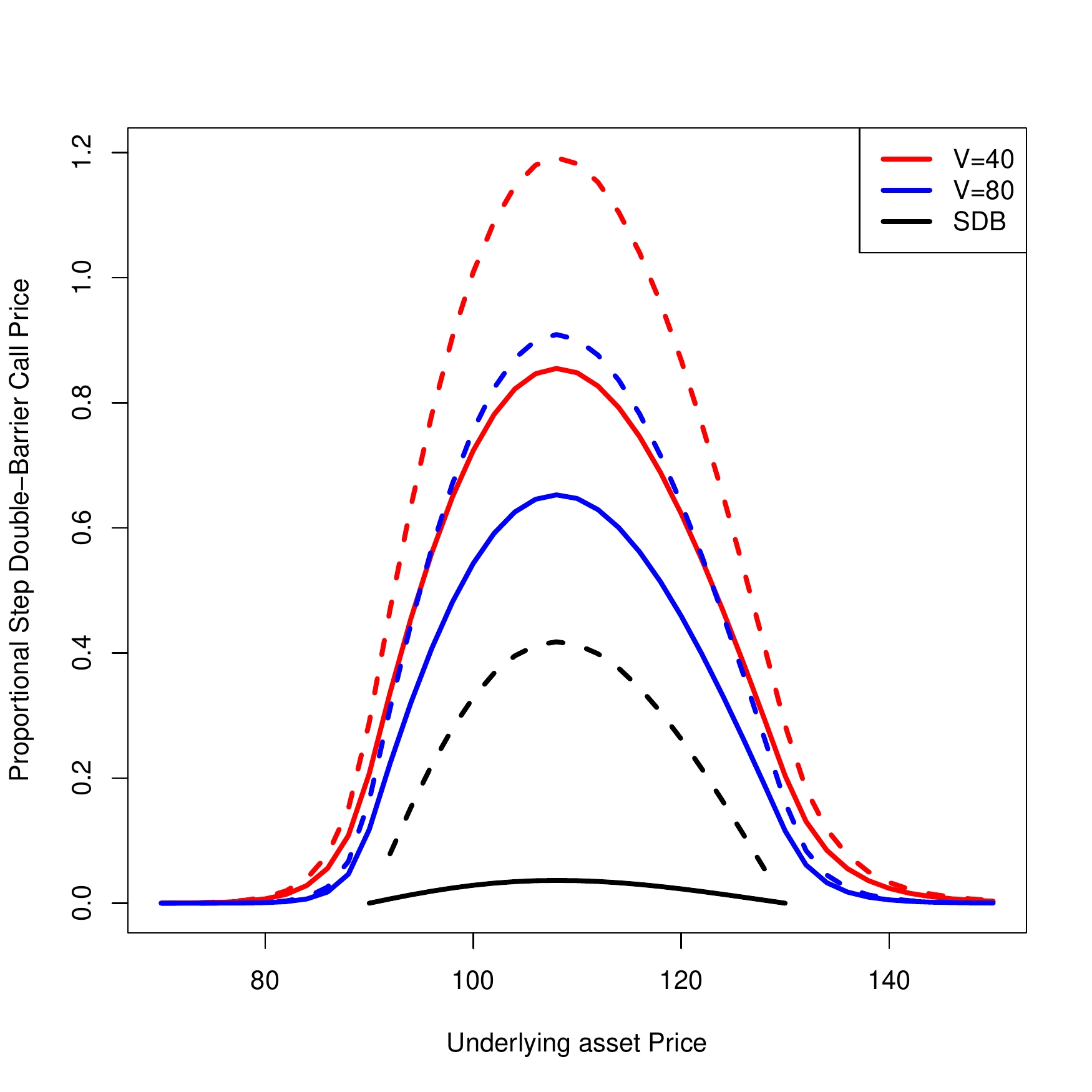}
\end{center}
\caption{Proportional double-barrier step call price as a function of  the underlying asset price with time-dependent interest rate $r(t)=0.04+0.01e^{-t}$ for different  potentials. The dashed lines indicate the corresponding fixed interest rate $r=0.05$ cases for comparison. Parameters: $a={\rm ln}100=4.5$, $b={\rm ln}130=4.867$, $K=100$, $\sigma=0.3$, $\tau=1$.}
\label{fig: Pricere}
\end{figure}

In Fig.~\ref{fig: Pricere}, we show the proportional double-barrier step call price  , with the time-dependent interest rate $r(t)=0.04+0.01e^{-t}$, as a function of underlying price. The dashed lines are corresponding to fixed interest rate $r=0.05$ cases for comparison. The option prices with an interest rate decreasing exponentially over time is lower than the one under fixed interest rate. 

\begin{figure}[!htbp]
\begin{center}
\includegraphics[width=0.49\linewidth]{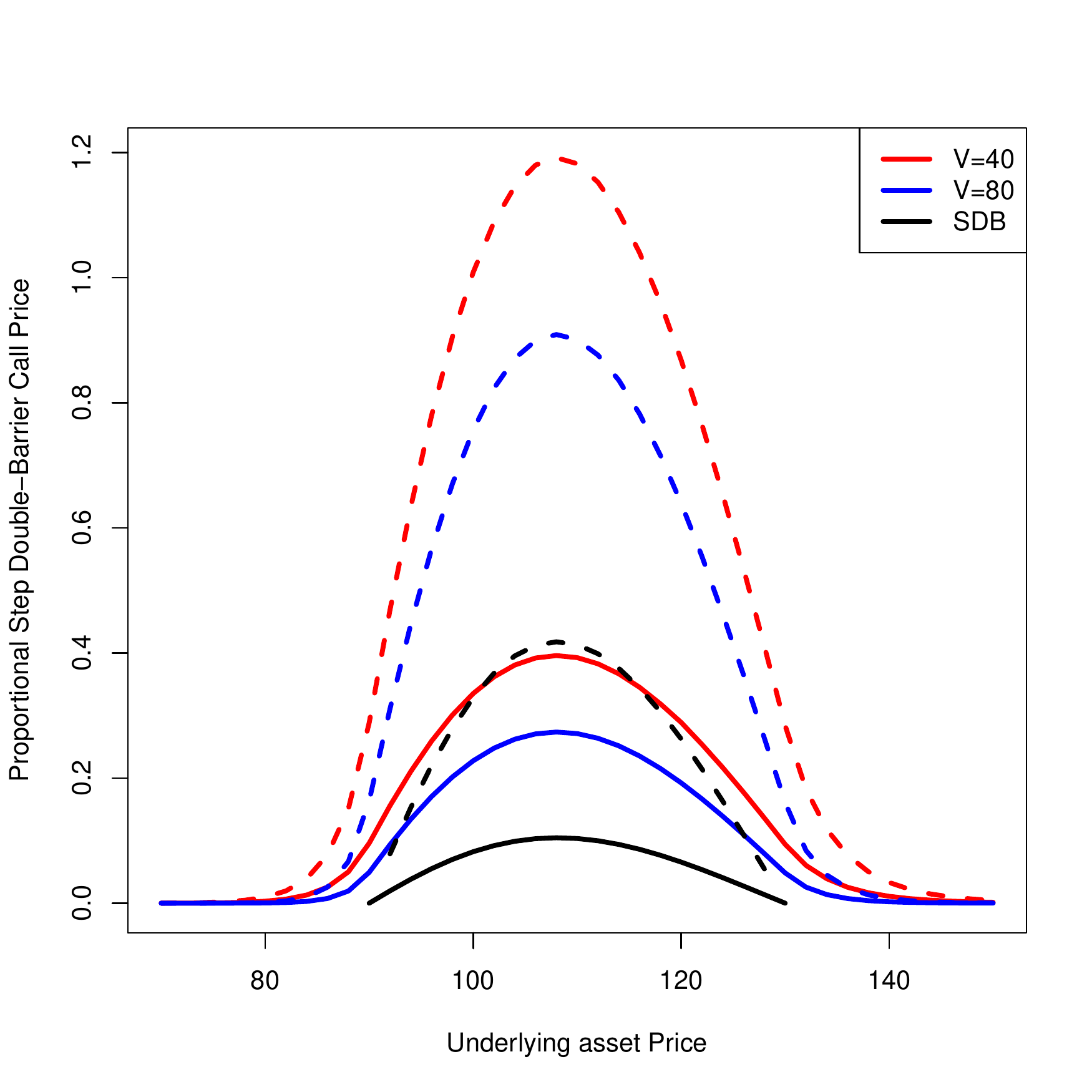}
\end{center}
\caption{Proportional double-barrier step call price as a function of  the underlying asset price with time-dependent volatility $\sigma(t)=0.3+0.05t$ for different  potentials. The dashed lines indicate the corresponding fixed volatility $\sigma=0.3$ cases for comparison. Parameters: $a={\rm ln}100=4.5$, $b={\rm ln}130=4.867$, $K=100$, $r=0.05$, $\tau=1$.}
\label{fig: Pricesig}
\end{figure}

In Fig.~\ref{fig: Pricesig}, we show the proportional double-barrier step call price  , with the time-dependent interest rate $\sigma(t)=0.3+0.05t$, as a function of underlying price. The dashed lines are corresponding to fixed volatility $\sigma=0.3$ cases for comparison. 

\section{Conclusion}

PDBS option price changing with time could be analogous to a particle moving in a finite square well. Considering the time dependent interest rate and volatility, infinite series summation and $\psi$ function should be used to acquire the analytical expressions for option price. We have discussed three different cases to demonstrate how to calculate the series. The path integral method could be generalized to price PDBS options with curved boundaries and American options.  
\appendix

\section{Path Integral Method for Proportional Double-Barrier Step Option Pricing}

The price changing of a proportional double-barrier step ({\rm PDBS}) option could be analogous to a particle moving in a symmetric square potential well with the potential
\begin{equation}\label{potentialPDBS}
V(x)=\left\{
\begin{aligned}
0 & , & a<x<b,\\
V_0 & , & x<a,\ x>b.
\end{aligned}
\right.
\end{equation}
for $x<a$ or $x>b$, the wave function decays with the increasing distances from the well, which is similar to an option touches a barrier and knocks out gradually. The Hamitonian for a double-barrier step option is~\cite{Baaquie}
\begin{equation}\label{nonhermitehamiPDBS}
H_{\rm PDBS}=-\frac{\sigma^2}{2}\frac{\partial^2}{\partial x^2}+\left(\frac{1}{2}\sigma^2-r\right)\frac{\partial}{\partial x}+r+V(x)
\end{equation}
which is a non-Hermitian Hamitonian. Considering the following transformation
\begin{equation}\label{hermitehamiPDBS}
H_{\rm PDBS}=e^{\alpha x} H_{\rm {eff}}e^{-\alpha x}=
e^{\alpha x}\left(-\frac{\sigma^2}{2}\frac{\partial^2}{\partial x^2}+\gamma\right)e^{-\alpha x}+V(x)
\end{equation}
where
\begin{equation}
\alpha=\frac{1}{\sigma^2}\left(\frac{\sigma^2}{2}-r\right),\\ 
\gamma=\frac{1}{2\sigma^2}\left(\frac{\sigma^2}{2}+r\right)^2
\end{equation}
and $H_{\rm {eff}}$ is a Hermitian Hamitonian which is considered as the symmetric square potential well Hamitonian. The stationary state Schr{\"o}dinger equation for option price is
\begin{equation}
    \left\{
\begin{aligned}\label{schrodingereqPDBS}
&-\frac{\sigma^2}{2}\frac{{\rm d}^2 \phi}{{\rm d}x^2}+\gamma \phi=E\phi  , & a<x<b,\\
&-\frac{\sigma^2}{2}\frac{{\rm d}^2 \phi}{{\rm d}x^2}+(\gamma+V_0) \phi=E\phi  , & x<a,\ x>b.
\end{aligned}
\right.
\end{equation}
where $\phi$ is the option price wave function, $E$ is corresponding to bound state energy levels in the potential well. (\ref{schrodingereqPDBS}) could be simplified into
\begin{equation}
    \left\{
\begin{aligned}\label{simpleschrodingerPDBS}
\frac{{\rm d}^2 \phi}{{\rm d}x^2}+k_1^2 \phi=0 & , & a<x<b,\\
\frac{{\rm d}^2 \phi}{{\rm d}x^2}-k_2^2 \phi=0 & , & x<a,\ x>b.
\end{aligned}
\right.
\end{equation}
where 
\begin{equation}\label{k1k2PDBS}
    k_1^2=\frac{2(E-\gamma)}{\sigma^2},\ \ k_2^2=\frac{2(V_0+\gamma-E)}{\sigma^2}
\end{equation}

The general solution for (\ref{simpleschrodingerPDBS}) is
\begin{equation}\label{general solution1}
\phi(x)=\left\{
\begin{aligned}
A_3\ e^{k_2(x-\frac{b+a}{2})} & , & x\leq a, \\
A_1 \sin(k_1x+\delta) & , & a< x \leq b, \\
A_2\ e^{-k_2(x-\frac{b+a}{2})} & , & x>b.
\end{aligned}
\right.
\end{equation}

Considering the continuity for both wave function and its derivative at $x=a$ and $x=b$, we have 
\begin{equation}\label{delta valuePDBS}
    \delta=\frac{\ell\pi}{2}-k_1\frac{b+a}{2},\ \ \ell=0, 1, 2,...
\end{equation}

According to different $\ell s$ in (\ref{delta valuePDBS}), (\ref{general solution1}) could be split into two parts

\begin{equation}\label{oddeqPDBS}
\phi_1(x)=\left\{
\begin{aligned}
&-A_2e^{k_2(x-\frac{b+a}{2})}  , & x\leq a, \\
&A_1 \sin k_1\left(x-\frac{b+a}{2}\right) , & a\leq x \leq b,\\ 
&A_2 e^{-k_2(x-\frac{b+a}{2})}  , & x>b.
\end{aligned}
\right.\ \ \ \ for\ \ \ell=0, 2, 4,...
\end{equation}
and
\begin{equation}\label{eveneqPDBS}
\phi_2(x)=\left\{
\begin{aligned}
&A_2 e^{k_2(x-\frac{b+a}{2})}  , & x\leq a, \\
&A_1 \cos k_1\left(x-\frac{b+a}{2}\right)  , & a\leq x \leq b, \\
&A_2 e^{-k_2(x-\frac{b+a}{2})}  , & x>b.
\end{aligned}
\right.\ \ \ \ for\ \ \ell=1, 3, 5,...
\end{equation}	
where 
\begin{equation}\label{A1A2PDBS}
    A_1=\sqrt{\frac{2k_2}{k_2(b-a)+2}},\ \ A_2=A_1\sin \bigg(k_1\frac{b-a}{2}\bigg)e^{k_2\frac{b-a}{2}}
\end{equation}
here the normalization condition has been used. Considering boundary conditions for (\ref{oddeqPDBS}) and (\ref{eveneqPDBS}) at $x=b$ respectively, we have
\begin{equation}\label{condition1PDBS}
    \cot k_1\frac{b-a}{2}=-\frac{k_2}{k_1}
    \end{equation}
    \begin{equation}\label{condition2PDBS}
    \tan k_1\frac{b-a}{2}=\frac{k_2}{k_1}
\end{equation}
let 
\begin{equation}\label{betaPDBS}
     \beta=\sqrt{k_1^2+k_2^2}=\frac{\sqrt{2V_0}}{\sigma}
\end{equation}
(\ref{condition1PDBS}) and (\ref{condition2PDBS}) could be combined into
\begin{equation}\label{conditionk1PDBS}
    k_{1n}\frac{b-a}{2}=\frac{n\pi}{2}-\arcsin{\frac{k_{1n}}{\beta}},\ \ n=1,2,3,...
\end{equation}
(\ref{conditionk1PDBS}) is the energy level equation. In general, there is no analytical solution for energy eigenvalues, yet numerical results for $k_{1n}$ could be obtained by Mathematica. In Table.~\ref{table: table1}, we show $k_{1n}$ for different $n$s at $V_0=55.7859$, $26.3401$, and $12.8233$ , which are corresponding to $d=0.8$, $0.9$ and $0.95$ in~\cite{Linetsky:2001}, respectively. The range of $k_{1n}$ could be obtained from (\ref{k1k2PDBS})
\begin{equation}
   k_{1n}<\sqrt{\frac{2(V_0-\gamma)}{\sigma^2}}
\end{equation}
\begin{table}

		\begin{center}
			\begin{tabular}{|p{2cm}|p{2cm}|p{2cm}|p{2cm}|p{2cm}|}
				\hline
				\makecell[c]{$V_0$} & \makecell[c]{$n$}  &    \makecell[c]{$k_{1n}$} & \makecell[c]{$n$} & \makecell[c]{$k_{1n}$}  \\
				\hline

			\multirow{3.1}{5.2em}{\makecell[c]{55.7859}}	&\makecell[c]{1}    &   \makecell[c]{7.38515}  & \makecell[c]{4} & \makecell[c]{28.8819} \\
			& \makecell[c]{2} & \makecell[c]{14.7215} & \makecell[c]{5}& \makecell[c]{34.4789}\\
			& \makecell[c]{3} & \makecell[c]{21.9384} & \makecell[c]{6}&\makecell[c]{42.1887}\\
				\hline
				
		  	    \multirow{2}{5.2em}{\makecell[c]{26.3401}}& \makecell[c]{1}     &   \makecell[c]{6.94708} &  \makecell[c]{3} & \makecell[c]{20.1958} \\
		  	    & \makecell[c]{2} &  \makecell[c]{13.7663} & \makecell[c]{4}&\makecell[c]{25.0978}\\
				\hline
				
			\makecell[c]{12.8233}&   \makecell[c]{1}    &   \makecell[c]{6.41782} &  \makecell[c]{2} & \makecell[c]{12.527}\\
				\hline
			\end{tabular}
		\end{center}
		\caption{$k_{1n}$ for different $n$s at $V_0=55.7859$, $26.3401$. Parameters: $\sigma=0.3$, $r=0.05$.}
		\label{table: table1}
	\end{table}
which restricts the number of energy levels is 5, 3 and 2 for $V_0=55.7859$, $26.3401$, and $12.8233$, respectively.

Now we calculate the pricing kernel of proportional double-barrier step option. a and b in (\ref{potentialPDBS}) could be considered as the lower and the upper barriers of the option. Let $\tau_1$ indicates the occupation time between the lower barrier a and the upper barrier b, and $\tau_2$ is the occupation time below the lower barrier a and above the upper barrier b. The pricing kernel is
\begin{equation}\begin{split}\label{pricingPDBS}
    P_{\rm PDBS}(x,x^\prime;\tau)&=\braket{x|e^{-\tau_1H_1-\tau_2H_2}|x^\prime}\\
    &=\int_{-\infty}^{+\infty}{\rm d}x^{\prime\prime}\braket{x|e^{-\tau_1 H_1}|x^{\prime\prime}}\braket{x^{\prime\prime}|e^{-\tau_2 H_2}|x^{\prime}}\\
    &=e^{-\tau\gamma}\int_{-\infty}^{+\infty}{\rm d}x^{\prime\prime}e^{\alpha(x-x^\prime)}\sum_n \sum_{n^\prime}e^{-\tau_1 E_{1n}-\tau_2 E_{2n^\prime}}\phi_n(x)\phi_n(x^{\prime\prime})\phi_{n^\prime}(x^{\prime\prime})\phi_{n^\prime}(x^\prime)
\end{split}\end{equation}
where $S_0=e^x$ is the initial price of the underlying asset, $S_T=e^{x^\prime}$ is the final price of the underlying asset. Using the orthonormalization condition
\begin{equation}
    \int_{-\infty}^{+\infty}{\rm d}x^{\prime\prime}\  \phi_n(x^{\prime\prime})\phi_{n^\prime}(x^{\prime\prime})=\delta_{nn^\prime}=\left\{
\begin{aligned}
0 & , & n \neq n^\prime,\\
1 & , & n=n^\prime .
\end{aligned}
\right.
\end{equation}
the pricing kernel (\ref{pricingPDBS}) is simplified into
\begin{equation}\begin{split}
    P_{\rm PDBS}(x,x^\prime;\tau)=e^{-\tau\gamma}e^{\alpha(x-x^\prime)}\sum_n e^{-\frac{1}{2}\tau\sigma^2k_{1n}^2}\phi_n(x)\phi_n(x^\prime)
\end{split}\end{equation}
where (\ref{k1k2PDBS}) is used, and the proportional double-barrier call price is 
\begin{equation}
    C_{\rm PDBS}(x;\tau)=\int_{{\rm ln}K}^{+\infty}{\rm d}x^\prime p_{\rm PDBS}(x,x^\prime;\tau) (e^{x^\prime}-K)
\end{equation}
where  
\begin{equation}\begin{split}
    H_1&=e^{\alpha x}\left(-\frac{\sigma^2}{2}\frac{\partial^2}{\partial x^2}+\gamma\right)e^{-\alpha x}\\
    H_2&=e^{\alpha x}\left(-\frac{\sigma^2}{2}\frac{\partial^2}{\partial x^2}+\gamma\right)e^{-\alpha x}+V_0
\end{split}\end{equation}
 $\phi_n(x)$ is the energy eigenstate in coordinate representation, $K$ is the exercise price, and $\tau=\tau_1+\tau_2$ is the time to expiration. Setting ${\rm ln}K\in (a,b)$, combining with the expressions of $\phi(x)$ in (\ref{oddeqPDBS}) and (\ref{eveneqPDBS}), the option price for $a<x<b$ could be denoted as 
 \begin{equation}
     C(x,\tau)\big|_{a<x<b}=\int_{{\rm ln}K}^b {\rm d}x^\prime P_{\rm PDBS}(x,x^\prime;\tau)\big|_{a<x<b,{\rm ln}K<x^\prime<b}(e^{x^\prime}-K)+\int_b^{+\infty}P_{\rm PDBS}(x,x^\prime;\tau)\big|_{a<x<b,x^\prime>b}(e^{x^\prime}-K)
 \end{equation}
 where
 \begin{equation}\begin{split}
     P_{\rm PDBS}(x,x^\prime;\tau)\big|_{a<x<b,{\rm ln}K<x^\prime<b}&=e^{-\tau\gamma}e^{\alpha(x-x^\prime)}(e^{x^\prime}-K)\times\\
     &\bigg(\sum_{n=2,4,6...}e^{-\frac{1}{2}\tau\sigma^2k_{1n}^2}A_1^2\sin\bigg[k_{1n}\bigg(x-\frac{b+a}{2}\bigg)\bigg]   
     \sin\bigg[k_{1n}\bigg(x^\prime-\frac{b+a}{2}\bigg)\bigg]+\\
     &\sum_{n=1,3,5...}e^{-\frac{1}{2}\tau\sigma^2k_{1n}^2}A_1^2\cos\bigg[k_{1n}\bigg(x-\frac{b+a}{2}\bigg)\bigg]
     \cos\bigg[k_{1n}\bigg(x^\prime-\frac{b+a}{2}\bigg)\bigg]\bigg)
     \end{split}\end{equation}
     \begin{equation}\begin{split}
    P_{\rm PDBS}(x,x^\prime;\tau)\big|_{a<x<b,x^\prime>b}&=e^{-\tau\gamma}e^{\alpha(x-x^\prime)}(e^{x^\prime}-K)\times\\
    &\bigg(\sum_{n=2,4,6...}e^{-\frac{1}{2}\tau\sigma^2k_{1n}^2}A_1A_2\sin\bigg[k_{1n}\bigg(x-\frac{b+a}{2}\bigg)\bigg]
     e^{-k_{2n}(x^\prime-\frac{b+a}{2})}+\\
     &\sum_{n=1,3,5...}e^{-\frac{1}{2}\tau\sigma^2k_{1n}^2}A_1A_2\cos\bigg[k_{1n}\bigg(x-\frac{b+a}{2}\bigg)\bigg]
     e^{-k_{2n}(x^\prime-\frac{b+a}{2})}\bigg)
 \end{split}\end{equation}
 
 Similarly, the option price for $x>a$ is
 \begin{equation}
 C(x,\tau)\big|_{x>a}=\int_{{\rm ln}K}^b {\rm d}x^\prime P_{\rm PDBS}(x,x^\prime;\tau)\big|_{x>a,{\rm ln}K<x^\prime<b}(e^{x^\prime}-K)+\int_b^{+\infty}P_{\rm PDBS}(x,x^\prime;\tau)\big|_{x>a,x^\prime>b}(e^{x^\prime}-K)
 \end{equation}
 where
 \begin{equation}\begin{split}
     P_{\rm PDBS}(x,x^\prime;\tau)\big|_{x>a,{\rm ln}K<x^\prime<b}&=e^{-\tau\gamma}e^{\alpha(x-x^\prime)}(e^{x^\prime}-K)\times\\
     &\bigg(\sum_{n=1,3,5...}e^{-\frac{1}{2}\tau\sigma^2k_{1n}^2}A_2A_1 e^{-k_{2n}(x-\frac{b+a}{2})}
     \sin\bigg[k_{1n}\bigg(x^\prime-\frac{b+a}{2}\bigg)\bigg]+\\
     &\sum_{n=2,4,6...}e^{-\frac{1}{2}\tau\sigma^2k_{1n}^2}A_2A_1 e^{-k_{2n}(x-\frac{b+a}{2})}
     \cos\bigg[k_{1n}\bigg(x^\prime-\frac{b+a}{2}\bigg)\bigg]\bigg)
 \end{split}\end{equation}
 \begin{equation}\begin{split}
     P_{\rm PDBS}(x,x^\prime;\tau)\big|_{x>a,x^\prime>b}&=e^{-\tau\gamma}e^{\alpha(x-x^\prime)}(e^{x^\prime}-K)\times\\
     &\bigg(\sum_{n=1,3,5...}e^{-\frac{1}{2}\tau\sigma^2k_{1n}^2}A_2^2 e^{-k_{2n}(x-\frac{b+a}{2})}
     e^{-k_{2n}(x^\prime-\frac{b+a}{2})}+\\
     &\sum_{n=2,4,6...}e^{-\frac{1}{2}\tau\sigma^2k_{1n}^2}A_2^2 e^{-k_{2n}(x-\frac{b+a}{2})}
     e^{-k_{2n}(x^\prime-\frac{b+a}{2})}\bigg)
 \end{split}\end{equation}

\end{document}